\documentstyle[aps,prb,preprint]{revtex}
\begin{document} 
\draft
\title{Magnetoquantum Oscillations in Mesoscopic Multi-Channel Rings}

\author{Yan Chen$^1$, Shi-Jie Xiong$^{2,1}$ and S. N. Evangelou$^{3}$}
\address{$^1$ Solid State Microstructure Laboratory 
 and Department of Physics,
       Nanjing University, Nanjing 210008, China \\ 
  $^2$China Center of Advanced Science and Technology (World Laboratory),
	P.O.Box 8730, Beijing 100080, China \\
  $^3$Department of Physics, University of Ioannina, Ioannina 45 110,
 Greece}

\date{\today}
\maketitle
\begin{abstract}

We obtain  exact analytical expressions for  the electronic
transport
through a multi-channel system, also with an applied 
magnetic field.
The geometrical structure of the electrodes is found 
to cause a
splitting of the conduction band into many subbands,
depending on the number and the length of the chains
and the  conductance approaches
zero when the chain number is sufficiently large, 
due to quantum interference.
In the presence of a  magnetic field  
a very  complicated  oscillatory behavior of the 
conductance is found
with a  very sensitive dependence on the number of chains
and their lengths, in a remarkable distinction 
from the usual  oscillations in
two-channel Aharonov-Bohm (AB) rings. In the multi-channel 
system  the
obtained oscillation patterns and their  periodicities 
depend  on the partitioning
of the magnetic flux in the areas enclosed by the electronic paths.
The present study may provide  a useful information
for quantum dots with a special configuration.

\end{abstract} 


\section{Introduction}   

Quantum transport through artificially fabricated nanostructures
has been extensively studied both experimentally and theoretically
during the past years \cite{1,2}.
The miniaturization of  quantum dots or wires has now 
reached a stage where devices  can be fabricated at sizes smaller
than the single particle electronic coherence length. 
In such mesoscopic systems the 
wavefunction maintains its phase coherence so that the
electron can travel coherently  through the sample.
The access to ``coherent transport'' is granted by the advances made 
in lithography techniques which have
opened a very rich field of theoretical
and experimental research concerning  quantum wires and quantum dots.
Electro-optical experiments in solid state devices could lead
to new switches which use the quantum wave nature of the electron.

Many interesting quantum effects can be also found in 
coupled nanostructures  where the electronic transport
is drastically affected by the phenomenom of
quantum interference.
Moreover, the application of a magnetic field, which  is
often used to probe the  properties of devices,  
can also induce characteristic changes in the phase coherence 
of the electronic wavefunctions \cite{3} which, in turn, 
give rise to  particular interference effects for the
electronic transport. In the  pioneering 
work of Aharonov and Bohm \cite{4} such an effect was
demonstrated via a thought experiment and it was shown 
that the conductance of a ring should
oscillate as a function of the magnetic flux 
threaded through it. Among the manifestations of the Aharonov
Bohm (AB) effect \cite{5,6,7,8,9}  usually  are  
the periodic magnetoresistance oscillations 
in normal metal rings and in electrostatically  defined
heterjunction rings.
The AB effect is  a result of the relative
phase shift between 
two electron beams enclosing a magnetic flux
$\phi$, where  the magnetic field causes
a $2\pi \frac{\phi}{\phi_{0}}$
change of the phase difference between the two 
arms of the ring. In this system the
magnetoresistance oscillations have period $\phi_{0}$,
which allows tuning of the phase  of a wavepacket with 
destructive and constructive interference in cycles.

Owing to the great variety of the possible configurations for 
quantum dots  it is of great interest to investigate the
change of the interference effect in AB rings when the
channel number  is greater than two. In this paper, we
study the electron transport 
properties of such a multi-chain structure with 
common leads attached at its ends. We show that 
it can provide  
many alternative options for tuning quantum interference effects
 in the electronic transport. In the multi-channel
structure an initial wave
splits up into  complementary waves $\psi_{1} \ldots \psi_{N}$,
where $N$ is the total number of chains involved.
These  waves propagate  independently  in every chain and are 
finally recombined  at the outgoing lead. 
Interference effects among the different waves can be 
observed from the behavior of the electrical resistance
calculated  between the two leads. We also show that if the
number of chains involved in the system is large enough most
of the states are reflected and only a few of them can propagate 
through the system.
This kind of ``blocking"  or ``localization" of the electron waves, 
which occurs via quantum interference effects caused 
by the geometrical structure, is observed
despite the absence of any disorder in the system.

In the presence of a magnetic field an electron moving 
around a loop will experience a phase change 
determined by the flux threaded through the loop. In 
a multi-chain system the phase changes are not the same
for different paths of propagation so that they can 
lead to   particular interference phenomena 
accompanied  by much more complicated conductance oscillations
than in the ordinary two-channel AB ring. We find that 
the pattern of these  magnetoquantum oscillations is very
sensitive to the number and the length distribution of the chains
involved in the structure.  Moreover, the 
oscillation patterns and their
periodicities are also very sensitive to the partitioning of the flux
among the areas enclosed by the paths. It should be 
pointed out that the  results 
obtained in this paper could be useful towards understanding
quantum dots  with a special configuration.

The structure of the paper with the exposition of our results
is as follows:
in section II we describe the studied structure and give  
analytic expressions for the electronic transport,
in section III we demonstrate different kinds of transport 
induced by special quantum interference effects 
with and without a magnetic field. 
The obtained  results 
are summarized and discussed in Section IV.

\section{Model and Formula}

We consider a ring of many chains
with two common leads at its ends,  threaded by a magnetic field which
produces a flux in every loop enclosed by two nearest-neighbor
chains. In addition we suppose that the multi-chain system is embedded
in an infinite perfectly conducting chain with
a left and a right part serving as the two electrodes. 
The configuration  is shown in Fig.1
and  the transport properties for non-interacting 
electrons in this system are studied via the 
tight binding Hamiltonian 
\begin{equation}
  H  = - t_0 \sum_{\scriptstyle \alpha =1}
 \: (  { c_{0}^{\dagger}\,  c_{\alpha ,1}} +
 e^{i\phi_{\alpha } }
  c_{\alpha ,N_{\alpha}}^{\dagger}  c_{s}  +
  \sum_{\scriptstyle i=1\atop}^{N_{\alpha} -1} 
 { c_{\alpha ,i}^{\dagger}\,  c_{\alpha ,i+1}}  +  
\text{ h.c. }),
\end{equation}
where $ c_{i,\alpha}$  $( c_{i,\alpha}^{\dagger})$ 
is the annihilation (creation) operator which annihilates (creates)  
an electron on the site $i$ of  chain $\alpha $,
$N_{\alpha}$ is the number of sites in the $\alpha $th chain 
(excluding the two nodes), $N$ is the total number of chains and
the two lead node sites are labelled by 0 and $s$.
The first two terms in the sum of Eq. (1) describe hopping
of the electrons between the ends of every chain 
and the two leads 0 and $s$.  Moreover, since
the chains are connected to  each other
only at their ends we can make a convenient choice of
the gauge for the vector potential to affect only
the phase of the wave functions  at the hopping bonds
between the right ends of each chain (site $N_{\alpha },
\alpha =1, 2, ..., N$) and the
right node $s$. In the Hamiltonian of Eq. (1) the
magnetic field is expressed by the second term in the  sum
where  the phase difference $\phi_{\alpha }-\phi_{\alpha -1}$
is proportional to the flux
$HW_{\alpha }$, $\alpha =2,3,...,N$, where $H$ is the strength 
of the field and $W_{\alpha }$  the area enclosed by
the $\alpha $ and the $(\alpha -1)$th chains. The phase 
of the first chain  is chosen to be zero  $\phi_1=0$
and the hopping strength for all the bonds $t_0=1$
is the energy unit used throughout the paper.

Our picture of the electronic transport in the system
consists of an electron wave  incident  from the source  
into the perfect chain, then ramified into the $N$ chains 
also experiencing different phase increments and
eventually recombined into one channel
at the output lead. Thus, an electronic beam incident from
the right should be partially transmitted and partially
reflected by the multi-chain system. 
In the site representation
the coefficients of the wave function at the 
left and the right parts of the pure chain can be
written as
\begin{equation}
\label{lead}
a_{j}=e^{-ikj}, \text{ for }  j\leq 0, 
a_{j}=Ae^{-ik(j-s)}+Re^{ik(j-s)}, \text{ for }  j\geq s ,
\end{equation}
where $k=\cos^{-1}(E/2)$ is the wave vector of a wave 
function with energy $E$, $A$ is the amplitude of the incident wave,
$R$ is the amplitude of the reflected wave and 
the wave function for all the bonds is normalized
so that the transmitted wave  amplitude  is unity. 
The transmission coefficient which
measures the transparancy of the system is, subsequently, 
defined as  $|t|^2=1/|A|^2$. 
The wave function coefficients in the $\alpha $th chain, by
including the left node 0, can be expressed as
a linear combination of the propagating and the 
reflected plane waves via
\begin{equation}
\label{chain}
a_{\alpha ,j}=A_{\alpha }e^{ikj}+R_{\alpha }e^{-ikj}, \text{ for }
0\leq j \leq N_{\alpha },
\end{equation}
where $a_{\alpha ,j}$ is the coefficient at the $j$th site of the 
$\alpha $th chain. The coefficient at the left lead node j=0  
from Eqs. (\ref{lead}) and (\ref{chain})
gives the relation
\begin{equation}
\label{eq1}
A_{\alpha}+R_{\alpha}=1.
\end{equation}
Similarly, we can calculate the coefficient at the right node j=$s$,
by including the flux-induced phase shift term, and a
comparison with Eq. (\ref{lead}) gives a second relation
\begin{equation}
\label{eq2}
(A_{\alpha}e^{ik(N_{\alpha }+1)}+R_{\alpha}e^{-ik(N_{\alpha }+1)})
e^{-i\phi_{\alpha}} =A+R.
\end{equation}
Therefore, from the two Eqs. (\ref{eq1}) and (\ref{eq2}) we can express 
the wave function coefficient of Eq. (3) for all chains $\alpha$ 
via
\begin{equation}
\label{01}
A_{\alpha }=-\frac{e^{-ik(N_{\alpha }+1)}-(A+R)e^{i\phi_{\alpha }}}
{2i\sin (k(N_{\alpha }+1))},
\end{equation}
\begin{equation}
\label{02}
R_{\alpha }=\frac{e^{ik(N_{\alpha }+1)}-(A+R)e^{i\phi_{\alpha }}}
{2i\sin (k(N_{\alpha }+1))},
\end{equation}
in terms of $A$ and $R$.

On the other hand from the Schr\"{o}dinger difference equations
at the two lead nodes 0 and $s$  we obtain
\begin{eqnarray}
E=e^{ik}+\sum_{\alpha }(A_{\alpha }e^{ik}+R_{\alpha }e^{-ik}),
\label{11}\\ 
\label{22}
E(A+R)=Ae^{-ik}+Re^{ik}+\sum_{\alpha }e^{-i\phi_{\alpha }}
(A_{\alpha }e^{ikN_{\alpha}}+R_{\alpha }e^{-ikN_{\alpha }}) 
\end{eqnarray}
and by a substitution of Eqs. (\ref{01}), (\ref{02})
into Eqs. (\ref{11}), (\ref{22}) eventually obtain
two equations for the $A$ and $R$. Finally, the  transmission
coefficient can be calculated from their solution, which gives
\begin{equation}
\label{sol}
|t|^2=\frac{4|f_0|^2\sin^2k}{|(c_0-e^{-ik})^2-|f_0|^2|^2},
\end{equation}
where
\[
f_0=\sum_{\alpha }\frac{e^{i\phi_1}e^{i(\phi_{\alpha }-\phi_1)}\sin k}
{\sin k(N_{\alpha }+1)},
\]
and
\[
c_0=\sum_{\alpha }\frac{\sin kN_{\alpha }}{\sin k(N_{\alpha }+1)},
\]
for arbitrary chain numbers $N_\alpha, \alpha = 1 ,2, ... , N$. 

Eq. (\ref{sol}) is the most important result of this paper,
which presents the general analytical expression for the 
transmission coefficient in a multi-channel ring. 
This expression can be further simplified for special geometries, for 
example, if the chain lengths are equal to
$N_1 = N_2 =... = N_N \equiv L$,  and the two nearest
neighbor chains  enclose a fixed area
$\phi = \phi_{\alpha }-\phi_{\alpha -1}$,  then 
$f_0$ and $c_0$  become
\begin{equation}
\label{sim1}
f_0=\frac{e^{i(N-1)\phi /2}\sin(N\phi /2)\sin k}{\sin(\phi /2)
\sin k(L+1)},
\end{equation}
\begin{equation}
\label{sim2}
c_0=\frac{N\sin kL}{\sin k(L+1)},
\end{equation}
and a simpler exact expression for the transmission
coefficient $|t(E)|^2$ can be obtained.
It can be seen from Eq. (\ref{sol}) that the magnetic field 
dependence is solely due to $|f_0|$ so that if the factors
$\frac{1}{W_2}$, $\frac{1}{W_2+W_3}$,...,
$\frac{1}{\sum_{\alpha =2}^NW_{\alpha }}$ have common multiples the 
oscillations of $|t(E)|^2$ as a function of a magnetic 
field have a period of $\nu\phi_0$, where $\nu$ is the smallest 
common multiple with $\phi_0$ the flux quantum. 
The electronic conductance can be also directly computed
from the transmission coefficient 
via the Landauer formula\cite{10}
\begin{equation}
\sigma(E)=\frac {|t(E)|^2}{ 1-|t(E)|^2},
\end{equation}
at a Fermi energy $E$.

\section{Quantum  oscillations for various multi-chain
configurations}

In this section we show our results 
associated with quantum interference effects in multi-chain
systems by application of Eq. (10). Our purpose is to
illustrate the relationship between
electronic wave transport in the multi-channel
system and the geometric ring structure, also in the 
presence of an external magnetic field.
								
\subsection{Equal chain system without a magnetic field}

In the absence of a magnetic field the replacement 
$\phi =0$ for $f_0$, $c_0$ is made in Eqs.
 (\ref{sim1}),(\ref{sim2}). 
The obtained results in this case  correspond  to
a similar model of a total number of $N$  thin wires joined
together at their two ends as it was  introduced by
Wang {\it et al}\cite{11} in their study of electronic transport
through a quantum cavity. These authors
have predicted that the total electron transmission can be 
simply expressed as a coherent sum of the transmission
coefficients obtained from every chain. 
Our results can give an even more complicated transmission 
behavior due to the geometrical structure of the electrodes.

In Fig.2 we plot the transmission coefficient versus the electronic
energy for such a multi-chain system made of  equal chains.
The pattern shown exhibits an interesting bridge-arc shape 
whose  curvature becomes larger and the blank region below
the arc smaller if the chain number increases, with  
some states still having high values of the 
transmission coefficient.  Thus,
if the chain number is large enough  most of the states are
reflected and only very few states can propagate
through the multi-channel system.
It must be emphasised that the 
blocking of the electron propagation at most
 energies is merely caused by quantum interference
due to the geometrical  structure involved.
It seems, however, a puzzle why such a rather 
symmetric geometry can give rise to a very 
complicated behavior of the outgoing wave. This is 
probably due to the fact that the  translation symmetry is  
broken at the two contacts,  which  leads  to
partial destructive interference of the electron waves.
Wang {\it et al} \cite{11} have also observed a partial
blocking of electron waves  in the propagation regime of a
quantum-wave-filter consisting of field-induced nanoscale
cavities and 1D wires by varying the electronic wavelength.
Our results could account for the reported
experimental behavior.

We have  also investigated the relationship between the
obtained features of the conducting spectrum and the
chain number for an equal chain multi-chain system. 
In Fig. 3 the  conduction band as a function of 
the number of  chains is illustrated and we
observe that by increasing the 
chain number the transmission pattern becomes more and more sparse
and the conduction band splits into several subbands.
If the number of chains becomes large enough  we find that
most of the states cannot 
propagate through the system, becoming
``blocked" or ``localized''. Thus, a
quantum interference effect  causes
the conduction band to become discrete 
in the absence of disorder and/or  interchain
couplings, except at
the connections of the system at the two end nodes.  
A relation between the 
conductance and the chain number can be extracted
from Fig.4,  where a monotonic drop of the
conductance  is seen when the number of chains
is increased. 
This is another indication of the trend shown by the
system for large chain numbers 
to become more ``insulating''.

\subsection{Magnetoconductance oscillations}

In Fig. 5(a) we show the characteristics of 
the transmission coefficient obtained in the
absence of a magnetic field, such as the 
bridge-arc shape already seen in Fig. 2(a), in order
to compare with the cases with an applied magnetic field
(Figs. 5(b), 5(c) and 5(d)). 
We find a remarkable change of the transmission 
in the latter case when
the areas  between neighboring chains
enclose equal  magnetic fluxes. In
Figs. 5(b), 5(c) and 5(d)  
the arc structure is no longer present and 
the transmission becomes more and more sparse
due to the  increase of the magnetic flux
through the system.

In Fig.6 we present the magnetic field
dependence of the transmission coefficient 
for equal chains with the same nearest-neighbor
path areas.  In this case 
the curves  show  periodic quantum-magnetic
oscillations governed by the field
dependence  which enters  $f_0$ via
Eqs. (11) and (12), finally leading to  $(N-1)\phi_0 $.
These findings share 
many similarities with the optical multi-slit 
interference patterns \cite{5,6} with 
main common feature the $N-1$ minima and the
$N-2$ subsidiary maxima between every two consecutive
principal maxima. However, the  obtained elctronic
transmission is more complicated when compared to the 
analogous optical case
due to the complexity of the denominator in 
our expression for $|t(E)|^2$.  
Moreover, from Fig.6   we can observe many points
of  zero transmission  which imply a
magnetic-field induced destructive interference effect.
  
Our results for a four-chain system with equal chains 
but  non-equal areas enclosed by every two
nearest-neighbor paths  are shown in Figs. 7 and 8.
It can be seen that the interference pattern and the period of
the magnetoscillation  vary, depending on 
the distribution of the magnetic flux between the closed paths.
In fig.7 we show  the electronic conductance versus the 
magnetic flux for a four-chain system made of equal chains
with magnetic flux periods 
(a) $3\phi_{0}$, (b) $5\phi_{0}$, (c) $4\phi_{0}$
and (d) $4\phi_{0}$. One can easily deduce the
relation between the oscillation period
and the distribution of the magnetic fluxes 
by noticing that the phase shift  for every chain
must be an integer times  $2\pi$.
If the magnetic flux distribution has a small deviation, 
the mageto-oscillation pattern and its period 
changes abruptly, as shown in Fig.8.
Without deviation, the spectrum has strict period $3\pi$ and
destructive interference occurs twice during this period. Introducing
the deviation, the spectrum has no strict periodicity
and the interference pattern changes aperiodically.

In Fig.9 we present  results for a four-chain system 
of both different chain lengths and non-equal areas
enclosed by two nearest-neighbor paths.  
From the realizations of Eq.(\ref{sol}) we find that the 
conductance  changes when the chain length varies 
because  of variations in both the numerator
and the denominator of Eq. (10).  
For a certain length distribution
we  observe a quasi-periodic pattern
close to about $1.5\phi_{0}$, but its
real period is $3\phi_{0}$ as in Fig. 9(d). Thus,
we  may conclude that even
a small variation of the chain lengths  causes
abrupt  changes in the conductance oscillation
patterns. It is, perhaps,  worth
mentioning that the sensitivity found could
provide  an opportunity  for the application of
the studied multi-chain structure to the
electronic device engineering.

\section{Discussion}

Quantum interference plays a central role in the quantum
physics of mesoscopic  systems. We have shown that for a 
multi-chain system  
an incident wave  splits into several chain
beams at the entrance and recombines at the exit.  Thus, the 
conduction band becomes discrete and the electronic
transport  properties are drastically modified by
the introduction of a ``localization" effect, despite 
the absence of any disorder. 
Moreover, in the presence of a magnetic flux
we obtain magneto-oscillations, which are much more 
complicated than these known in the usual AB rings.
In the AB effect a magnetic field is  
threaded through the center of a ring so that the
electrons passing via each of the two chains
experience different phase shifts. If the 
magnetic field is varied one can modulate
the phase  and produce conductance oscillations in the
wave transport from one terminal to the other.

The magnetic field dependence of the
electrical conductance also shows an oscillating  
behavior very different from the AB ring effect,
since the multi-chain system exhibits more complicated interference 
effects determined by the phase shifts in  
the various propagation paths. Each phase shift is caused by
both the electronic momentum and  the magnetic flux,
so that momentum  variations and changes in the chain  
lengths  as well as variations in the 
distribution of the magnetic fluxes can modify 
the  interference  pattern.  Electron wave propagation through
a multi-chain system pierced by a magnetic field has also
an interesting analogy  with optical 
interference phenomena  by many slits. In both phenomena
$N-2$  subsidiary maxima 
and $N-1$   minima between two consecutive principal
maxima occur. Of course,
between each pair of minima 
a subsidiary maximum exists \cite{12}, as it is confirmed by
our  numerical calculations. It must be pointed 
that our results are also relevant to the case of 
Andreev scattering  \cite{13}
which occurs in normal-superconductor interfaces. It turns 
out that if the right hand side periodic chain attached to
the dot structure is
replaced by a clean superconducting wire our
obtained results of the transmission coefficient 
for the normal-dot-normal can be also 
used for finding the transmission through the
normal-dot-superconductor system. This can be achieved via 
Beenaker's formula \cite{13} for 
the conductance 
$G= (2 e^{2}/h) {\frac {2 |t|^{4}}{(2-|t|^{2})^{2}}}$
expressed from the 
transmission of the
non--superconducting dot part  only. 
If we use the obtained $|t|^{2}$ from
Eq. (10)  extra  doubling of the 
periodicities should occur  for the 
dot-superconductor interface.

In summary, we have systematically studied the electronic
properties of a multi-chain system connected at its two ends. 
A recursion method was employed and an exact analytic  expression
for the  electronic conductance  was presented.
Many interesting features in the transmission coefficient and the 
magnetoconductance were also shown for various configurations:  
1) The  geometrical structure of the electrodes is found
to cause a discreteness of the conduction band,
which eventually affects remarkably the transport properties
leading to a kind of ``localization" in the absence of disorder. 
2) We find various magneto-oscillation periodicities and interference
patterns, by  varying the distribution of the relative magnetic
flux through the structure, and also
abrupt changes in the plot of the conductance 
versus the magnetic-flux 
if the length distribution of the system 
is modulated, which is useful to distingush even  slight 
chain length variations. 
3) The studied system can be also used to probe the
distribution of the magnetic field since the obtained
interference patterns are  very sensitive 
to the distribution of the
magnetic flux among  neighbouring closed paths. 

\newpage
\vspace{0.15in}
 
\begin{center} 
 
\vspace{0.3in} 
 
\topmargin=-5mm 
  
\baselineskip=7mm 
\begin{large} 
 
{\bf ACKNOWLEDGEMENT} 
\end{large}  
\end{center}  
     
This work was supported by the National Natural Science  
Foundation of China, a $\Pi. EN. E. \Delta.$ grant of the 
Greek Secretariat of Science and Technology, HCM 
and TMR programs of the E.U. We also like to thank Prof.
Colin Lambert for pointing out the relevance of our
results for the normal-superconducting interface.

\newpage 
\vspace{0.02in} 
\begin{figure}

Figure 1. The considered multi-chain system with the left and right 
nodes indicated by 0 and $s$, respectively. The total number
of chains is $N=5$ and the number of sites in the $\alpha$th
chain is $N_\alpha, \alpha= 1, 2, ... ,N$ without
counting of the nodes 0  and $s$. 
 
\vspace{0.12in} 
 
Figure 2. The transmission coefficient  as a function of the
electronic energy for a N-chain system of equal chain lengths
(a) $N_\alpha = 100, \alpha= 1, 2, ...,N$ and
(b) $N_\alpha = 1000, \alpha= 1, 2, ...,N$. The chain numbers
involved in each case are: 
(1) $N=2$, (2) $N=4$,  (3) $N=10$, (4) $N=40$  and (5) $N=80$.

\vspace{0.12in} 
 
Figure 3. The  conduction band  vs.
the chain number $N$ for a multi-chain system with chain
lengths $N_\alpha=5000, \alpha= 1, 2, ... ,N$. 
A conduction band is
defined as being non-zero at the energy values 
where the corresponding transmission coefficient 
is higher than 0.1.

\vspace{0.12in} 
 
Figure 4. The conductance $\sigma(E)$ as a function of the
number of chains $N$  for a system with equal chains  
at a  Fermi energy  $E=1.0$, in the absence of 
a magnetic field. 
 
\vspace{0.12in} 
 
Figure 5. A comparison of the transmission coefficient with
and without a magnetic field.  The structure consists of $N=4$
channels of lengths  $N_\alpha = 2000, \alpha= 1, 2, 3, 4$
and the magnetic flux threaded in the 
system is: (a) 0, (b) 0.1, (c) 0.5 and (d) 2.0.

\vspace{0.12in} 
 
Figure 6.  The electronic transmission vs. the magnetic flux  
for  a multi-chain system with equal 
chain lengths $N_\alpha =2000, \alpha =1,2,...,N$
  and fixed electron energy 
$E=1.1$.  The unit of the magnetic flux is the 
flux quantum $\phi_0=1$ and the chain numbers are:
(a) $N=2$, (b) $N=3$, (c) $N=4$, d) $N=5$ and e) $N=9$. 
 
\vspace{0.12in} 
 
Figure 7. The electronic conductance vs. the  magnetic
flux  for a multi-chain system ($N=4$) 
with equal chain lengths
$N_\alpha = 2000, \alpha= 1, 2, 3, 4$
and  electron energy  fixed at $E=1.1$, with 
the magnetic flux quantum $\phi_0=1$. 
(a) $\phi_2-\phi_{1}$=$\phi_3-\phi_{2}$=$\phi_4-\phi_{3}$,  
(b) $\phi_2-\phi_{1}$=$\phi_3-\phi_{2}$=$\frac{1}{2}
             (\phi_4-\phi_{3})$
(c)  $\phi_2-\phi_{1}$=3$\phi_3-\phi_{2}$ and $\phi_4-\phi_{3}$=0,  
(d)  $\phi_2-\phi_{1}$=$\phi_3-\phi_{2}$=
            $2(\phi_4-\phi_{3})$. 
 
\vspace{0.12in} 
 
Figure 8. The electronic conductance vs. the magnetic
flux for a multi-chain system ($N=4$) with equal chain lengths
$N_\alpha = 2000, \alpha= 1, 2, 3, 4$
and  electron energy fixed at  $E=1.1$, with 
the magnetic flux quantum $\phi_0=1$:
(a) $\phi_2-\phi_{1}$=$\phi_3-\phi_{2}$=
            $\phi_4-\phi_{3}$,
(b)  $0.99(\phi_2-\phi_{1})$=$\phi_3-\phi_{2}$=
            $1.01(\phi_4-\phi_{3})$,
(c)  $0.98(\phi_2-\phi_{1})$=$\phi_3-\phi_{2}$=$1.01
            (\phi_4-\phi_{3})$.

\vspace{0.12in} 

Figure 9.The electronic conductance vs. the magnetic
flux for a multi-chain system  ($N=4$) 
 with almost equal chain lengths
$N_\alpha , \alpha= 1, 2, 3, 4$
and  electron energy fixed at  $E=1.1$, with the magnetic
flux quantum $\phi_0=1$:
(a) the chain lengths are (a) $N_\alpha =2000, \alpha =1,2,3,4$,
(b) $N_1 =2000, N_2 =2002, N_3 =2004, N_4 =2006$,
(c) $N_1 =2000, N_2 =2010, N_3 =2020, N_4 =2030$
(d) $N_1 =2000, N_2 =2100, N_3 =2200, N_4 =2300$.

\end{figure}
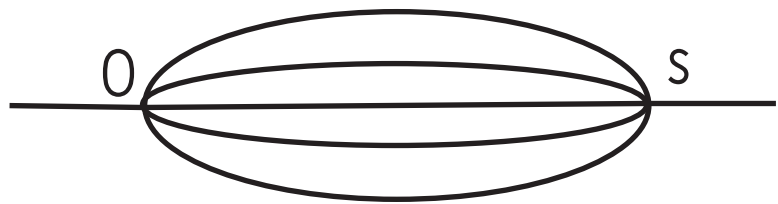
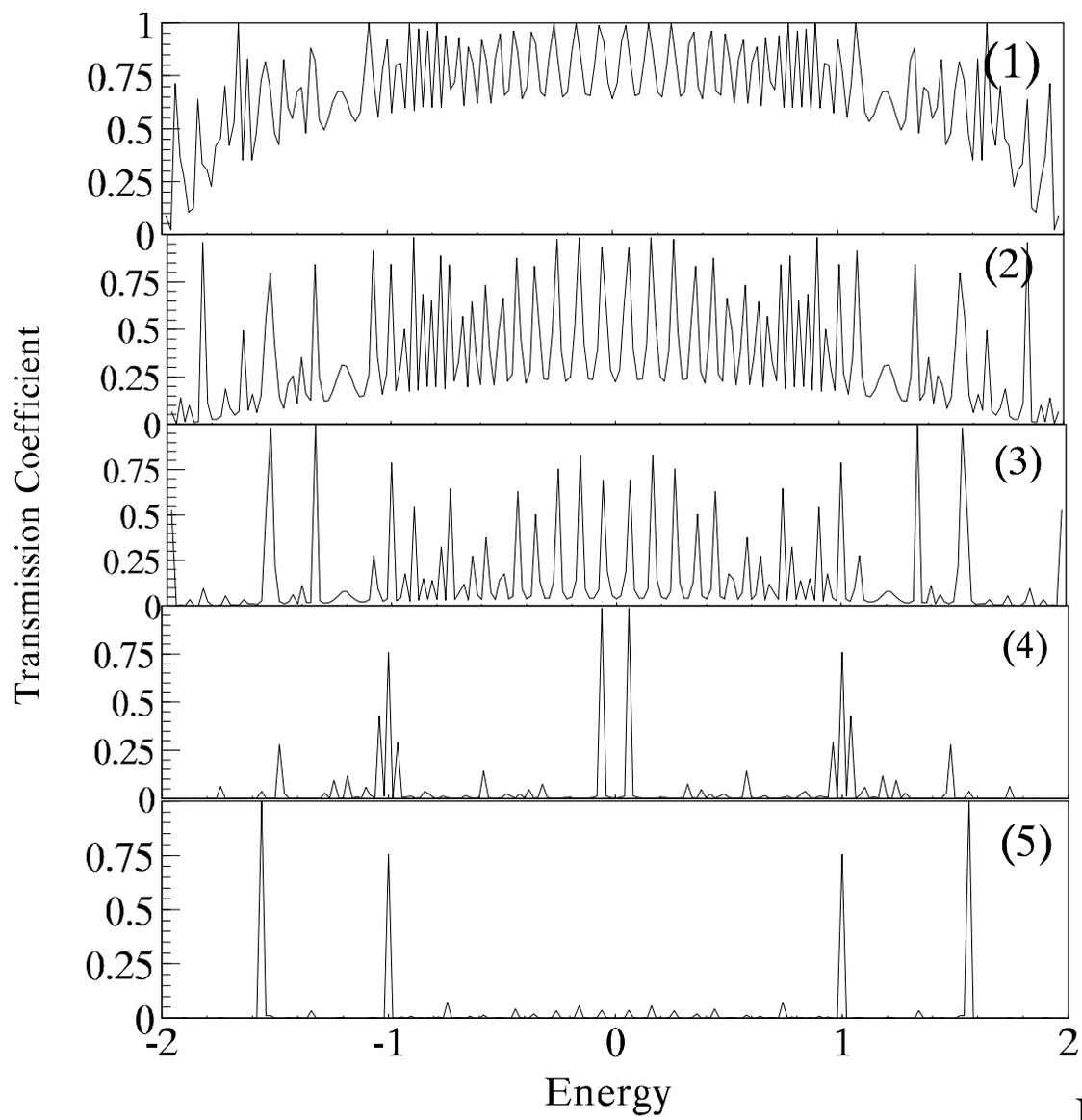
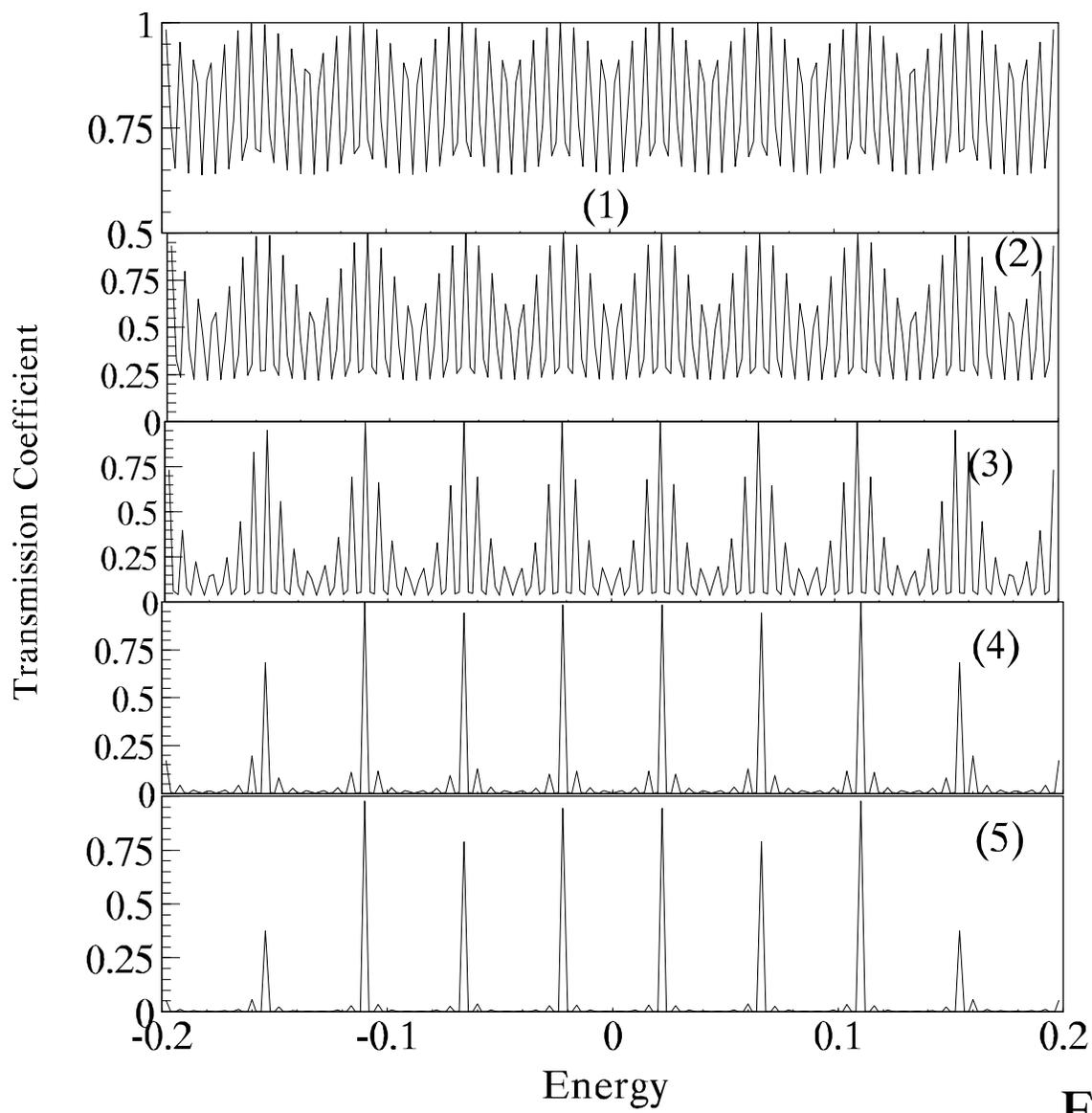
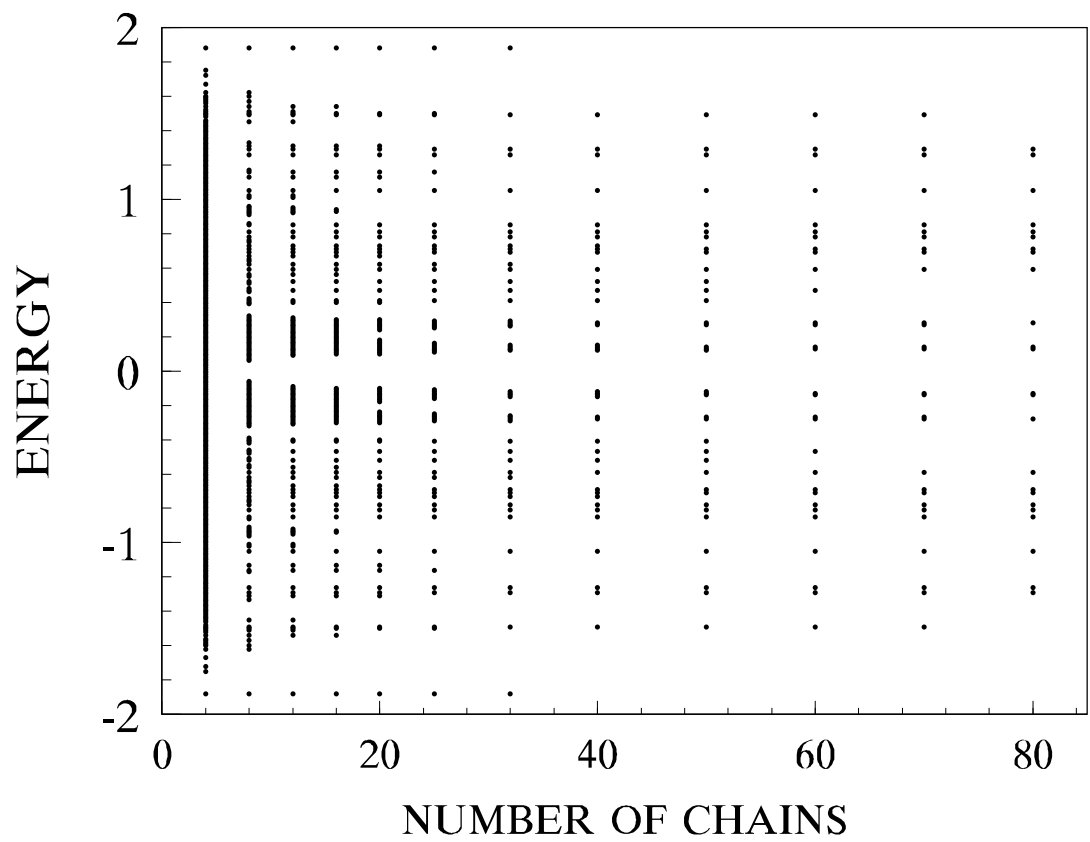
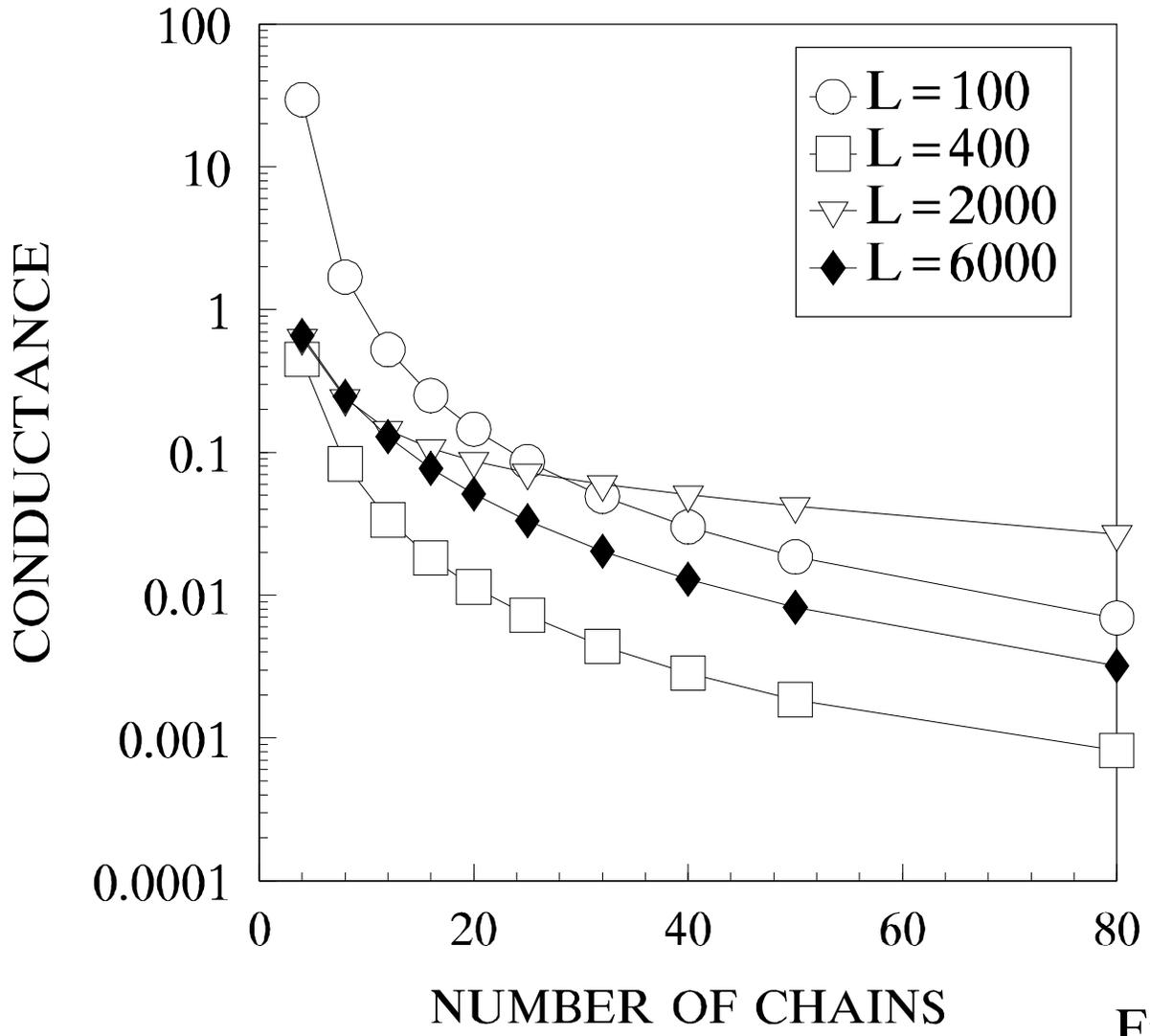
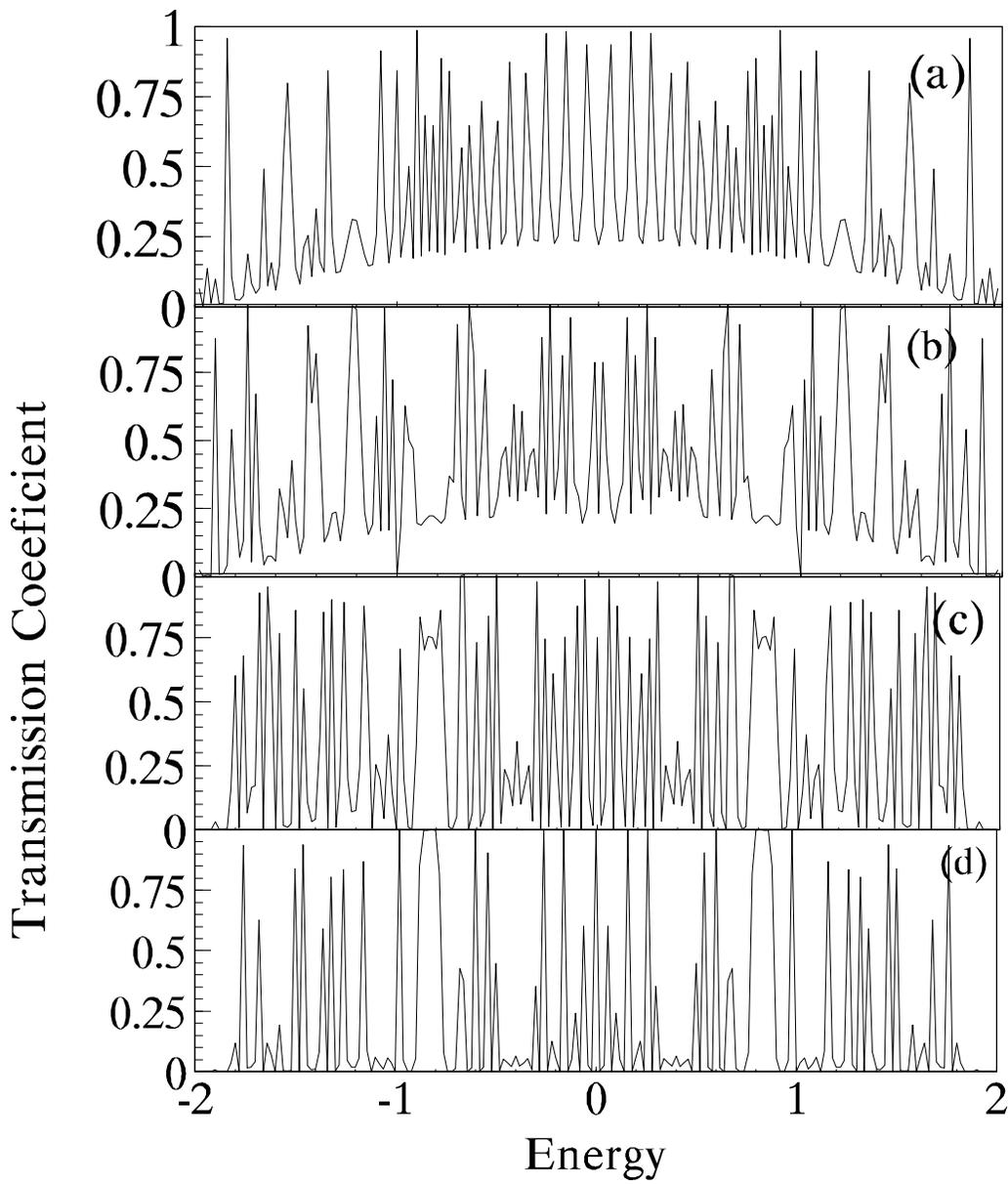
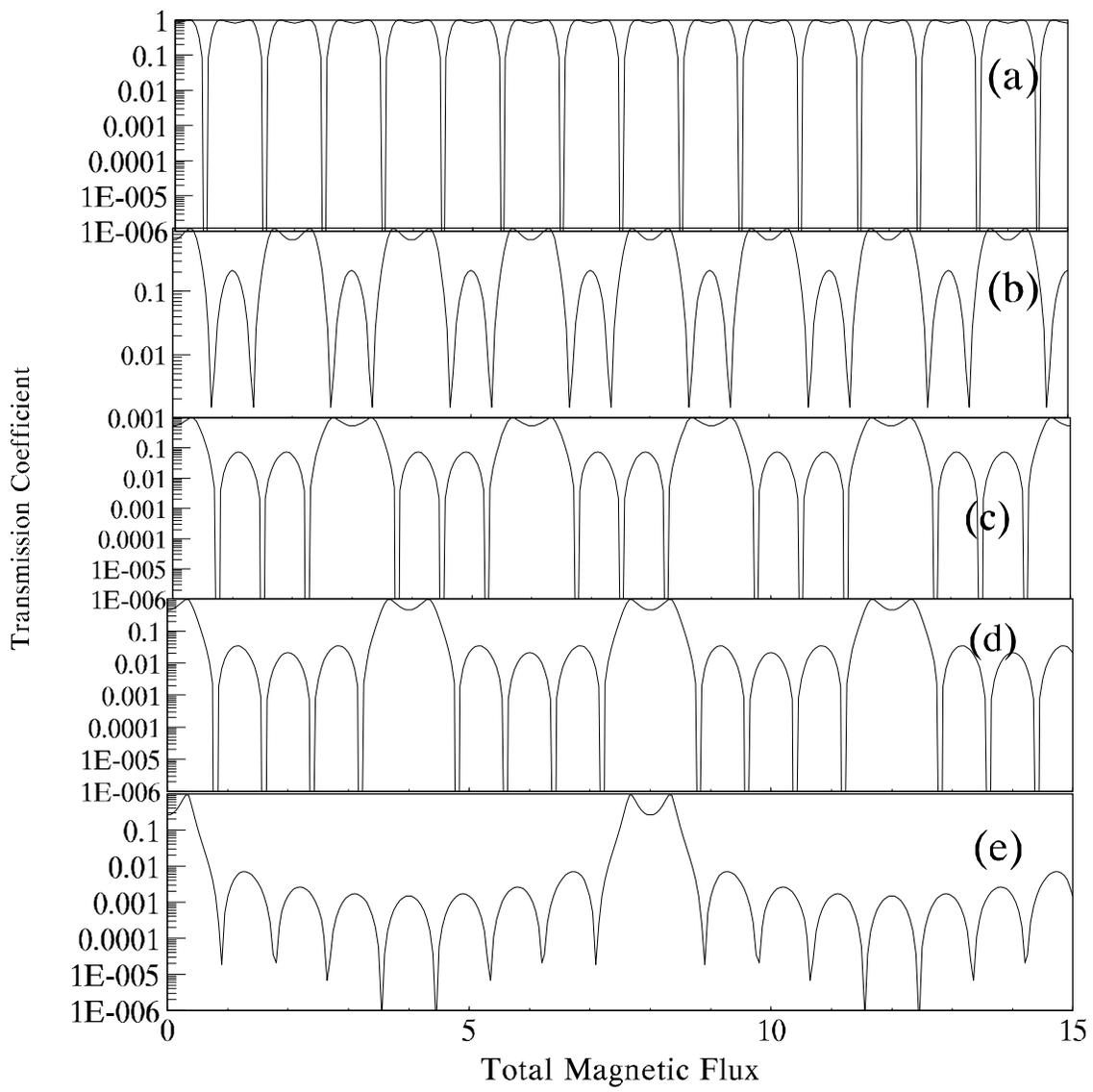
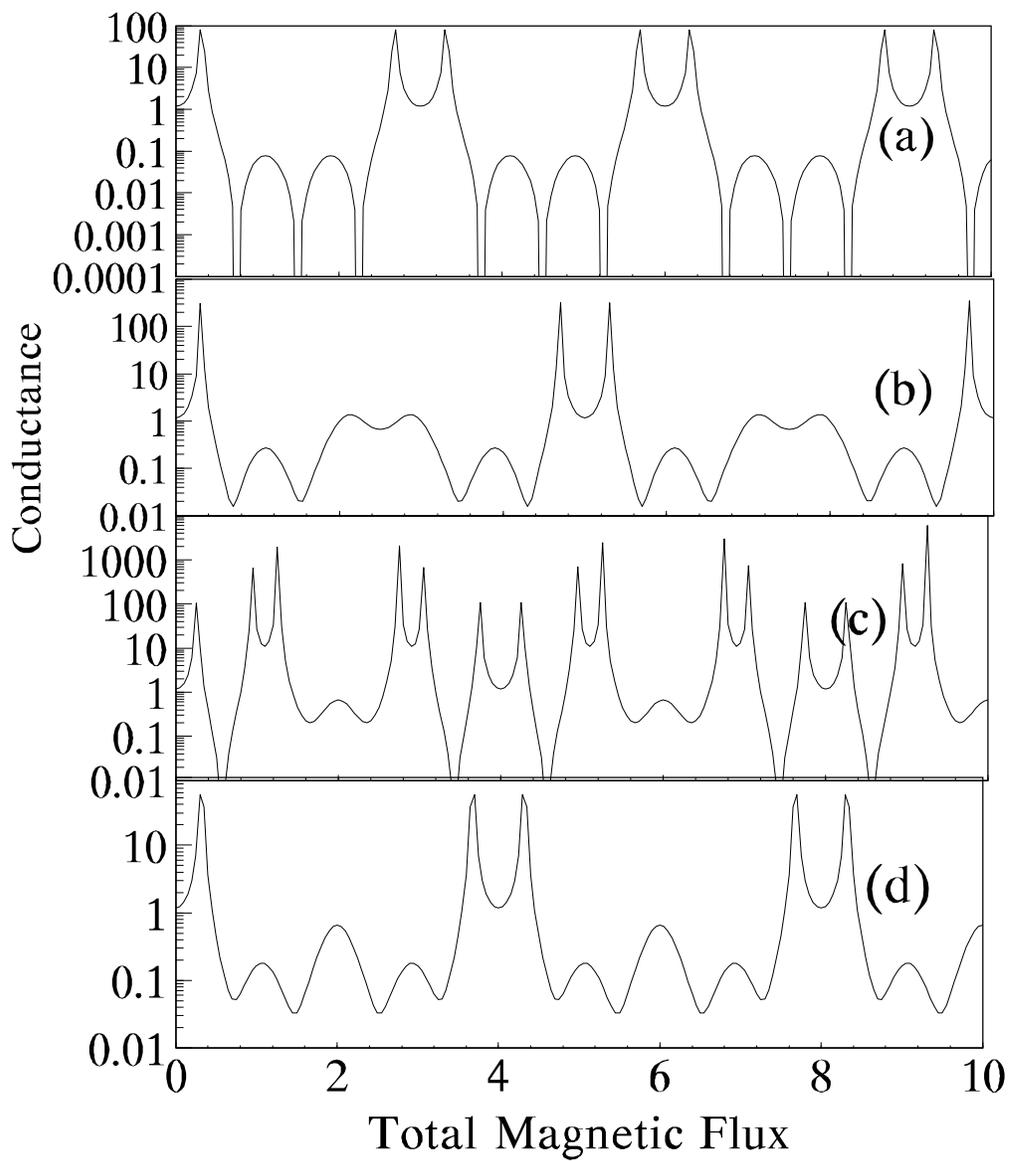
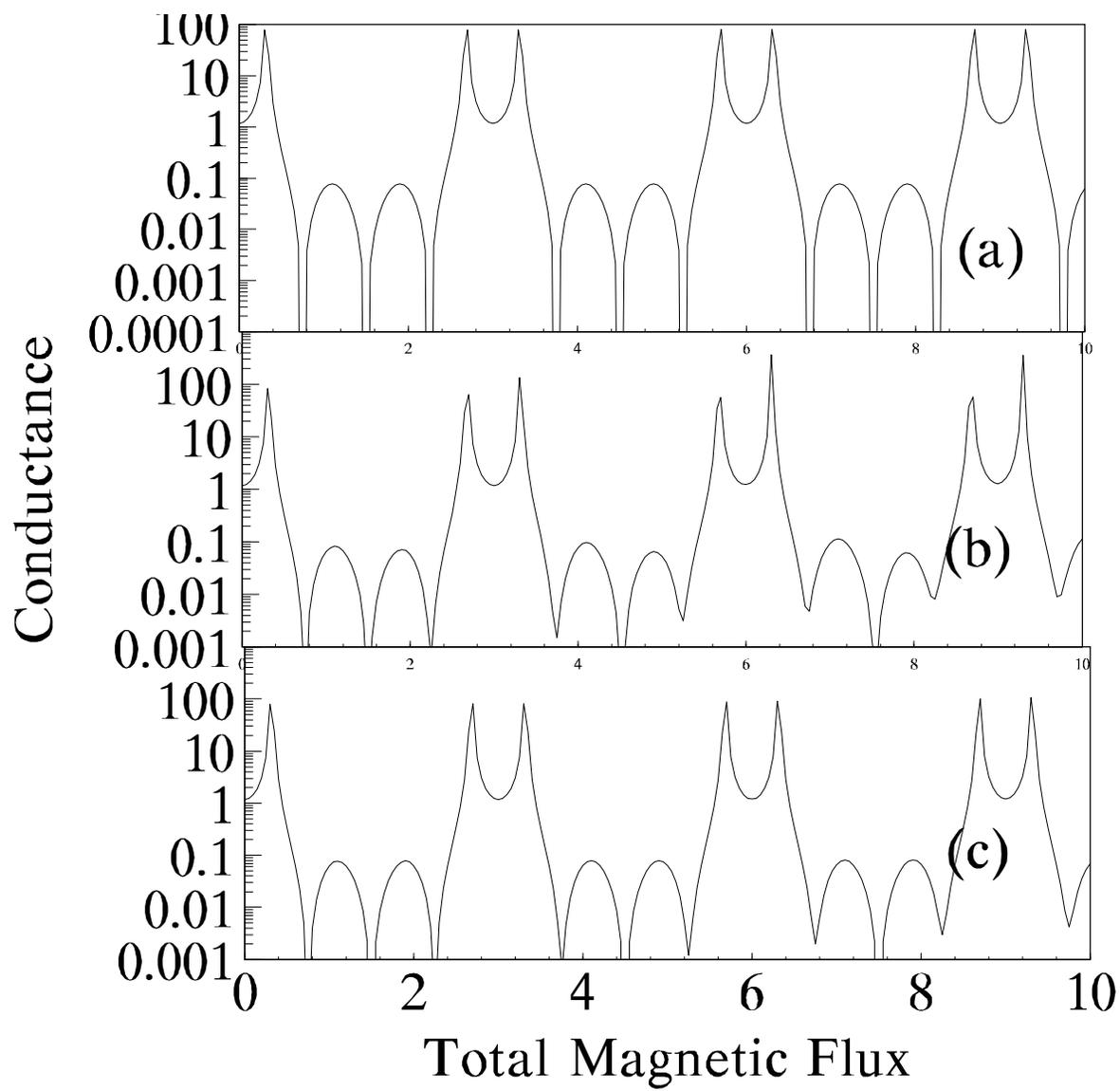
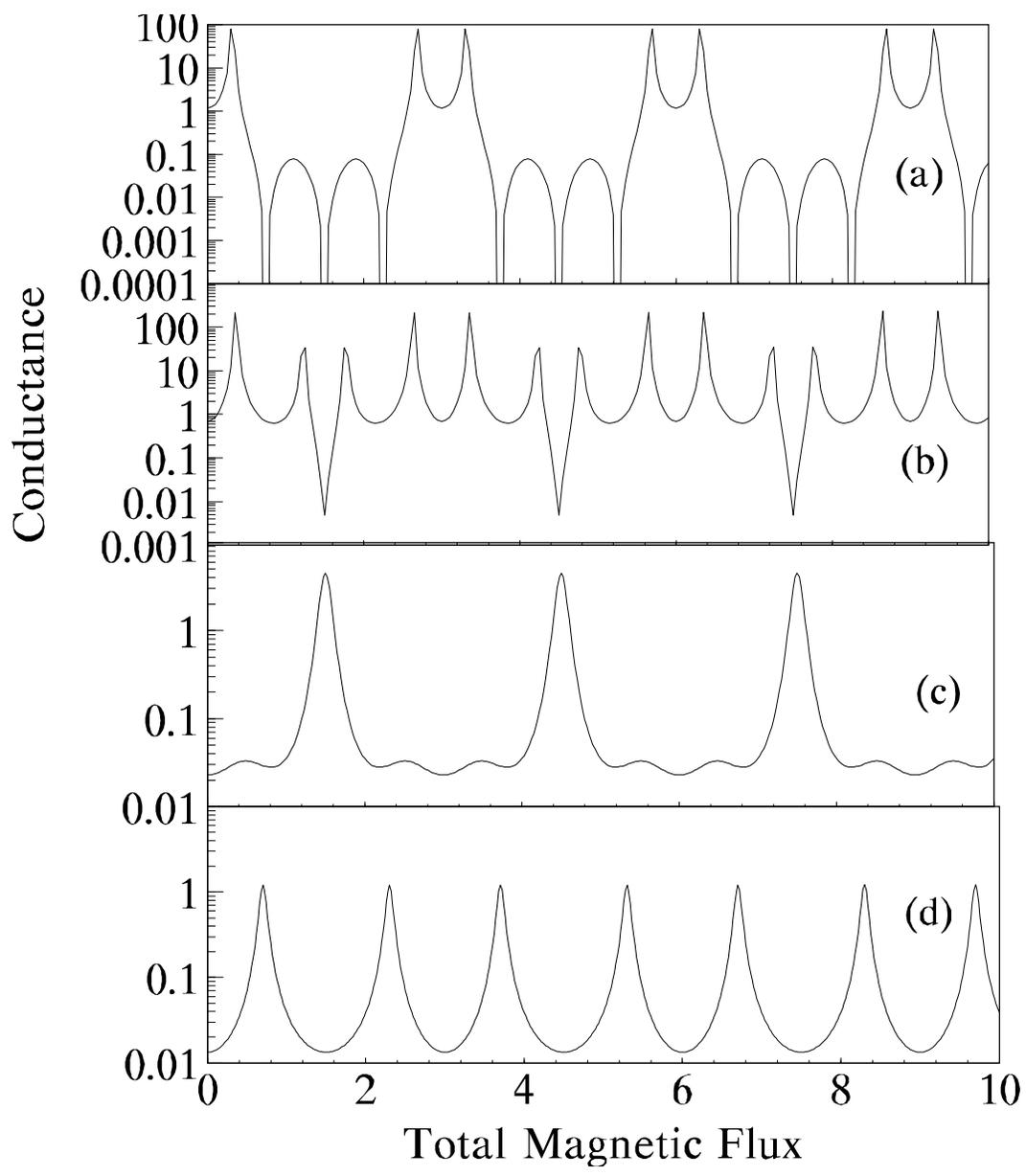
\end{document}